# The State-of-Practice in Requirements Elicitation: An Extended Interview Study at 12 Companies

Cristina Palomares[1], Xavier Franch[1], Carme Quer[1], Panagiota Chatzipetrou[2], Lidia López[1], Tony Gorschek[3]

1. Software and Service Engineering Group (GESSI)
   Universitat Politècnica de Catalunya (UPC), Barcelona (Catalonia, Spain)
   Jordi Girona num.1-3, Omega building, room S206
   {cpalomares,franch,cquer,llopez}@essi.upc.edu
2. Department of Informatics, School of Business,
   Örebro University,
   Örebro, Sweden
   panagiota.chatzipetrou@oru.se
3. Software Research Engineering Lab (SERL)
   Blekinge Institute of Technology,
   Karlskrona, Sweden
   tony.gorschek@bth.se

*Abstract*

***Context.*** Requirements engineering remains a discipline that is faced with a large number of challenges, including the implementation of a *requirements elicitation process* in industry. Although several proposals have been suggested by researchers and academics, little is known of the practices that are actually followed in industry.

***Objective.*** We investigate the state-of-practice with respect to *requirements elicitation*, by closely examining practitioners' current practices. To this aim, we focus on the techniques that are used in industry, the roles that requirements elicitation involves, and the challenges that the requirements elicitation process is faced with.

***Method.*** We conducted an interview-based survey study involving 24 practitioners from 12 different Swedish IT companies. We recorded the interviews and analyzed these recordings by using quantitative and qualitative methods.

***Results.*** We found that *group interaction techniques*, including meetings and workshops, are the most popular type of elicitation techniques that are employed by the practitioners, except in the case of small projects. Additionally, practitioners tend to use a variety of elicitation techniques in each project. We noted that customers are frequently involved in the elicitation process, except in the case of market-driven organizations. Technical staff (for example, developers and architects) are more frequently involved in the elicitation process compared to the involvement of business- or strategic staff. Finally, we identified a number of challenges with respect to stakeholders. These challenges include difficulties in understanding and prioritizing their needs. Further, it was noted that *requirements instability* (i.e., caused by changing needs or priorities) was a predominant challenge. These observations need to be interpreted in the context of the study.

***Conclusion.*** The relevant observations regarding the survey participants' experiences should be of interest to the industry; experiences that should be analyzed in the practitioners' context. Researchers may find evidence for the use of academic results in practice, thereby inspiring future theoretical work, as well as further empirical studies in the same area.

# 1 Introduction

*Requirements elicitation* (RE) is typically seen as the first step in the requirements engineering process (Nuseibeh & Easterbrook, 2000a). This step refers to the activities that are undertaken to reveal the requirements of a system that is to be built or a problem that is to be solved (Sommerville & Kotonya, 1998). Elicitation is not merely a matter of transcribing exactly what users say (Wiegers & Beatty, 2013). Instead, elicitation should be understood as *the search for*, *the gathering of*, and *the consolidation of* a project's requirements. This is no easy task, as testified by industry reports with respect to the magnitude of this challenge. For example, it is claimed that "80% of products lose money due to wrongly set product objectives that result in building products that customers do not need" (Stallabaum & Ly, 2016). Furthermore, the quality of this process is critical to the building of successful solutions, because the detection of errors at the initial stages of the product development process can save considerable time and money (see Nuseibeh & Easterbrook, 2000a; Procaccino et al., 2002; van Lamsweeerde, 2009). Nuseibeh et al. (2000b) highlight the fact that requirements do not exist somewhere 'out there', merely waiting to be collected, but, rather, elicitation techniques are crucial to the proper investigation, identification, and understanding of the users' needs.

In the challenging arena described above, several researchers have approached requirements elicitation from different perspectives. For example, Carrizo et al. (2014) and Pacheco et al. (2018) both present literature reviews on elicitation techniques, including questionnaires, interviews, analysis of existing documentation, group elicitation techniques (e.g., focus groups, workshops), prototyping, model-driven techniques (e.g., based on scenarios, KAOS, or *i\**), cognitive techniques (e.g., protocol analysis, laddering, card sorting), contextual techniques and creativity techniques (e.g., brainstorming and role playing). The literature reviews conducted by Dieste and Juristo (2011) and Ambreen et al. (2018) also report on empirical studies (primarily case studies and experiments) that assess the effectiveness of some of these elicitation techniques and models. Other recent work has taken into consideration a number of current challenges and recent opportunities in this area, even addressing crowd-based RE (Groen et al., 2017) and data-driven RE (Maalej et al., 2016, 2019).

The above-mentioned research notwithstanding, the RE community has published only a few studies that explore the way in which requirements elicitation is actually conducted in industry. This relative lack of research calls for additional empirical studies to be conducted; studies which can serve to further inform us about the way practitioners work. They should address research questions as: What methods are used in current state-of-practice? What roles are involved? What are the challenges faced by practitioners during this activity that remain to be solved? These questions have motivated the work that is presented in this paper, and should be regarded as corresponding to our claim that the research and development of new ways of working with elicitation techniques should be informed by the state-of-practice in industry.

We report on the results of an interview-based empirical study involving 12 IT companies and 24 experienced senior practitioners. This is done in order to present an understanding of how requirements elicitation practices are performed in industry. The paper directly addresses three research questions regarding the techniques used, the roles involved, and the challenges faced in requirements elicitation. We analyze these results and compare them with findings presented by existing studies in the field.

The paper is structured as follows: Section 2 presents other work that is related to the present study in the form of a summary of previous empirical studies that report on industrial practices with respect to requirements elicitation. Section 3 describes the research methodology that is used in this study. Section 4 informs the reader of the results of the interview study, and Section 5 presents an analysis of these results. Finally, Section 6 consists of our conclusions and recommendations for future work.

## 2 Related Work

In this section, we present the most relevant empirical research reports on industrial practices in the form of empirical studies on requirements elicitation published since 2010. Some of these studies consist of general studies on RE that include some results on elicitation (see Section 2.1), while the rest are more focused on elicitation only (see Section 2.2).

### 2.1 *Empirical RE studies on industrial practices including results on elicitation*

In Table 1, below, we summarize several empirical studies on industry-based requirements engineering, including results on elicitation.

Elahi et al. (2011) is oriented towards the study of security RE practices in China. One of the main conclusions of this paper is that security requirements are not often explicitly elicited and documented in the early stages of the development process, but, instead, are mostly considered during the implementation phase. Another conclusion is that requirements analysts elicit security requirements isolated from known vulnerabilities, but there is a widespread use of security standards that usually stem from compliance requirements and audit processes, and common attacks occurred in the past are generally considered.

Raatikainen et al. (2011), in their study of RE practices in nuclear industry, report on the challenge of achieving efficient communication during requirements elicitation and, more specifically, how to communicate efficiently with stakeholders with different backgrounds and across organizational borders.

Bjarnason et al. (2011) report the existence of communication gaps that may cause problems during elicitation. When the gaps are between requirement engineers and stakeholders, they may lead to miss vital requirements. On the other hand, a weak vision of the overall goal may lead to a wrong decision on which requirements to include in a project.

Berntsson Svensson et al. (2012) examine a number of specific challenges associated with the selection, tradeoff, and management of quality requirements (QRs). Regarding elicitation, they report on the practice of eliciting interdependencies among requirements. The results indicate that few of the companies actually manage to a large extent to effectively elicit, analyze, and document interdependencies. They also report that QRs were quantified by most of the participants in the study (from the 22 participants, 8 participants "always" quantified, whilst 12 participants "sometimes" quantified).

Hiisilä et al. (2015) highlight a number of challenges that are related to the elicitation of requirements. These include: (i) the scoping and planning of the project; (ii) the difficulty in arriving at a common understanding amongst the stakeholders and the supplier(s); (iii) the difficulty in reaching an agreement amongst the stakeholders on the final needs of the system; and (iv) the lack of stakeholder cooperation in the early RE phases.

*Table 1. RE empirical studies on industrial practices, including results on the elicitation of requirements*

|  | ***Study*** | ***Topics with Relevant Results on Elicitation*** |
|---|---|---|
| (Elahi et al. 2011) | • *Goal:* To come to an understanding of security RE Chinese practices<br>• *Type of study:* Questionnaire-based survey<br>• *Population:* 374 subjects from 237 companies of different domains and sizes<br>• *Country:* China | • *Security RE practices* |
| (Raatikainen et al. 2011) | • *Goal:* To study the state-of-practice in RE<br>• *Type of study:* Interviews<br>• *Population:* 7 subjects from 4 companies dealing with safety-related automation systems in nuclear energy<br>• *Country:* Finland | • *Elicitation challenge* |
| (Bjarnason et al. 2011) | • *Goal:* To understand the causes and effects of communication gaps<br>• *Type of study:* Interviews<br>• *Population:* 9 subjects that deal with requirements and development of a large market-driven software company on the embedded systems domain<br>• *Country:* Not stated | • *Communication gaps existence*<br>• *Causes of the communication gaps*<br>• *Effects of the communication gaps* |
| (Berntsson et al. 2012) | • *Goal:* To investigate the elicitation, analysis and negotiation, management, and handling of quality requirements in industry<br>• *Type of study:* Interviews<br>• *Population:* 22 subjects (11 product managers, 11 project leaders) from 11 software companies<br>• *Country:* Sweden | • *Quantification of requirements* |
| (Hiisilä et al. 2015) | • *Goal:* To investigate the challenges faced by a customer company's RE process in an outsourced development environment<br>• *Type of study:* Interviews; Project analysis; Workshops to validate the results<br>• *Population:* 17 subjects in interviews, analysis of 15 projects and 5 workshops, all of them from an insurance company<br>• *Country:* Finland | • *Elicitation challenges* |
| (Kassab et al. 2015) | • *Goal:* To investigate the evolution of requirements engineering practices during a ten-year period.<br>• *Type of study:* Three online questionnaire-based surveys<br>• *Population first run:* 194 subjects<br>• *Population second run:* 93 subjects<br>• *Population third run:* 247 subjects<br>• *Country:* USA | • *Evolution of elicitation techniques* |
| (Wagner et al. 2019) | • *Goal:* To study the state-of-practice in RE<br>• *Type of study:* Questionnaire-based surveys<br>• *Population first run:* 58 companies (one participant per company)<br>• *Country first run:* Germany<br>• *Population second run:* 228 companies (one participant per company)<br>• *Countries second run:* 10 countries | • *Elicitation techniques*<br>• *RE standards*<br>• *Requirements classification*<br>• *Elicitation challenges* |
| Malviya et al. (2017) | • *Goal:* To identify the different kinds of questions that business analysts and requirements engineers are interested in asking to support requirements-related tasks<br>• *Type of study:* Questionnaire-based survey<br>• *Population:* 29 valid survey responses (from 40 responses, 101 persons were contacted by e-mail and the survey was advertised in Linkedin groups)<br>• *Country:* Worldwide | • *Elicitation techniques* |
| Liebel et al. (2018) | • *Goal:* To identify specific problems/challenges in automotive RE with respect to communication and organization structure<br>• *Type of study first run:* Interviews<br>• *Population first run:* 14 interviews (2 companies)<br>• *Type of study second run:* Surveys<br>• *Population second run:* 31 valid survey responses<br>• *Country:* 4 countries | • *Automotive RE problems*<br>• *Automotive RE challenges* |

| Alsaqaf et al (2019) | • *Goal:* To identify challenging situations in agile quality requirements engineering and industrial practices so as to be able to mitigate the effects of such situations<br>• *Type of study:* Semi-structured open-ended interviews<br>• *Population first run:* 17 interviews<br>• *Country first run:* Netherlands | • *Elicitation challenges*<br>• *Elicitation practices* |
|---|---|---|

Kassab et al. (2015) present the results of a questionnaire-based survey conducted in 2013 on the state-of-practice of RE, that replicates two surveys conducted in 2003, 2008, and analizes the evolution occurred. They observe that the use of scenarios as elicitation technique decreased during the timeframe of their study, compared to interviews, which increased considerably in their use. If only agile projects were considered, the decrease was even more dramatic, whilst user stories emerged in this context.

Malviya et al. (2017) conducted a questionnaire-based survey to elicit information about the questions that are asked by requirements engineers. The authors classified the collected questions in 9 different purposes, being one of them elicitation, with several sub-purposes as, for instance, the identification of different types of requirements, the identification of related requirements, the discovery of sources of requirements that could be reused and the management of requirements. The type of artefacts that participants reported to address elicitation questions were, for instance, metrics, models, previous projects, process steps, requirement types or stakeholders.

Liebel et al. (2018) present the results of 14 semi-structured interviews in which one of the researchers' goals was to identify the problems and challenges in automotive RE related to the communication structure of the two companies. Seven problems were identified and validated by answers to a questionnaire given by 31 practitioners in the automotive domain, but 'a lack of product knowledge' and 'insufficient resources for understanding requirements' were considered to be related to requirements elicitation.

Alsaqaf et al. (2019) identify, by means of 17 semi-structured interviews, a number of challenging situations in agile QR engineering. With respect to current industry practices that are used to mitigate the impact of the challenges that agile QR engineering is faced with, one of the identified challenges was 'the identification of QRs'. The practices that are used to overcome this challenge are (i) to establish components and preparation teams and (ii) to reserve part of the sprints for important QRs.

To date, the most comprehensive empirical study in the RE field is the NaPiRE[1] initiative. It proposes a survey be conducted periodically so as to investigate the state-of-practice and the current problems that RE practitioners are faced with. The NaPiRE has been conducted twice so far; one NaPiRE survey was conducted in Germany (Méndez et al. 2015) and the second survey was conducted in a group of 10 countries (Wagner et al. 2019). Regarding the state-of-practice of elicitation techniques, the results of these surveys indicate that the most frequently used elicitation techniques consist of (i) interviews, (ii) facilitated meetings including workshops, and (iii) prototyping (Wagner et al., 2019). With respect to the challenges that RE is faced with, the NaPiRE surveys conducted in Austria and Brazil (Kalinowski et al. 2015) report that the most frequently mentioned challenges are: (i) the existence of incomplete and/or hidden requirements; (ii) moving targets (i.e., rapid changes in the requirements); (iii) time boxing; (iv) the difficulties faced by stakeholders in separating requirements from previous solutions; (v) requirements that are too abstract and allow for various interpretations; and (vi) existence of communication flaws between project team and customer.

---

[1] http://www.re-survey.org/

## 2.2 Empirical RE studies reporting on industrial practices that are focused on elicitation aspects

In this section, we report on a number of studies that are related to industrial practices regarding the elicitation of requirements. A summary overview of these studies can be found in Table 2 below.

Liu et al. (2010) studied the (then) state-of-practice with respect to elicitation and specification of requirements in Chinese companies by means of a survey that was answered by 377 people. Liu et al. (2010) determine that (i) the more used elicitation techniques are face-to-face meetings, (ii) more than 50% of people also use other communication media and (iii) other used techniques are rapid prototyping and reference to similar systems. Regarding problems in the elicitation due to changes in the requirements once signed the contract with customers, only the 14% follow what is stated in the contract, whilst 80% try to negotiate with the customer on the changes introduced. Also, they report differences among elicitations depending on the type of company: meanwhile, in multi-national corporations there are people specialized in RE tasks and they use multiple elicitation techniques, in governmental companies there are not dedicated RE positions, therefore it is usual to have communication gaps between customer and development team, and users may revoke confirmed requirements even in later stages of projects.

Bjarnason et al. (2011) report the existence of communication gaps that emerge during requirements elicitation, primarily caused by practitioners not having a clear vision of the overall goal. This gives rise to a situation with low levels of motivation with respect to contributing to requirements work, incorrect and unclear requirements, and general quality challenges with system requirement specifications.

Todoran et al. (2013) examine the question of whether (and if so, how) the elicitation process that takes place in cloud systems differs from the process used in traditional systems. They further enquire whether current techniques suffice. The results of their study show that interviews, questionnaires, analysis of existing documentation, surveys, and prototyping are the most popular and frequently applied techniques.

Hadar et al. (2014) present an empirical study where they examine the perceived- and actual effects of prior domain knowledge on requirements elicitation via interviews. The first part of the study was run with students, but the second part involved industrial participants, hence its inclusion in this related work. Hadar et al.'s (2014) results indicate that domain knowledge affects elicitation via interviews in two main aspects: (i) communication with the customers and (ii) understanding their needs.

Sethia et al. (2014) deduce a causal relationship between requirements elicitation issues and project performance. The challenges related to the elicitation of requirements that are presented in the results of their study are organized into three groups: (i) scope, which are those challenges related to the need of having different requirements for different stakeholders; (ii) volatility, which is related to the extent of changes that the requirements undergo during the project life cycle; and (iii) understanding challenges which are related to the degree of requirements understanding required in the elicitation process.

Manzoor et al. (2018) find that cloud providers use traditional elicitation techniques and that most cloud service providers use more than one elicitation technique depending on the consumer. Up to 40% of the companies which are served by cloud providers are highly dissatisfied with the elicitation techniques that are used however. The authors further deduce that the elicitation techniques that are used are not sufficient and can easily lead to customer dissatisfaction.

*Table 2. RE empirical studies on industrial practices that are focused on the elicitation of requirements*

|  | ***Study*** | ***Topics with Relevant Results*** |
|---|---|---|
| (Liu et al. 2010) | • *Goal:* To know the state-of-practice of RE in Chinese companies<br>• *Scope:* Elicitation and Specification of requirements<br>• *Type of study:* Survey<br>• *Population:* 377 subjects from 237 companies of different domains and sizes<br>• *Country:* China | • *Elicitation practices*<br>• *Elicitation challenges* |
| (Todoran et al. 2013) | • *Goal:* To investigate the state of practice of the adoption and implementation of existing elicitation techniques in cloud systems development<br>• *Scope:* Elicitation of requirements<br>• *Type of study:* Interviews<br>• *Population:* 26 subjects with a good overview of the elicitation process and related company needs of 19 companies that are cloud providers<br>• *Countries:* India, USA, UK | • *Elicitation techniques* |
| (Hadar et al. 2014) | • *Goal:* To examine the perceived and actual effects of prior domain knowledge on requirements elicitation via interviews<br>• *Scope:* Elicitation of requirements<br>• *Type of study:* Experiments and interviews to validate results<br>• *Population:* In two experiments, 31 and 38 participants, respectively, who were enrolled on a requirements university course; 5 RE professionals participated in the interviews<br>• *Country:* Not stated | • *Aspects affected by the level of domain knowledge in elicitation via interviews* |
| (Sethia et al. 2014) | • *Goal:* To establish and validate an empirical model that is used to study the behavior between requirements elicitation issues and project performance<br>• *Scope:* Elicitation of requirements<br>• *Type of study:* Online survey<br>• *Population:* 203 subjects involved in RE in different companies that are focused on different domains<br>• *Country:* Not stated | • *Elicitation challenges* |
| (Manzoor et al. 2018) | • *Goal:* To identify the elicitation methods used by cloud services providers, and the satisfaction ratings of the end user<br>• *Scope:* Elicitation of requirements<br>• *Type of study first run:* Online questionnaire<br>• *Population first run:* Cloud provider employees<br>• *Type of study first run:* In-depth interviews<br>• *Population first run:* Cloud provider employees<br>• *Country:* Pakistan | • *Elicitation techniques used by cloud services providers* |

## 3 Research Methodology

The investigation that is presented in this paper was carried out by using a qualitative research approach which included in-depth semi-structured interviews (Robson, 2002). Qualitative research aims to investigate and understand phenomena within their real-life context. This approach is useful when the purpose of an investigation is to explore an area of interest or to improve our understanding of a phenomenon (Robson, 2002) (Runeson & Höst, 2009). Since the purpose of our study is to investigate the use of RE practices in companies, we considered this research approach to be the most appropriate. The study focuses on several aspects of RE. Given the large amount of information that was collected for this purpose, we decided to divide the reporting of our results across several papers. The present paper reports on the results for the elicitation part of this study. In addition, we have written the full protocol as a separate document that is available as supplementary material (protocol.pdf). Consequently, in this paper, we report just the most relevant facts on the protocol, and refer the interested reader to this separate document for additional details.

## 3.1 Summary of the protocol

***Research Questions.*** The research questions (RQs) that are addressed in this paper are listed in **Table 3**. The first RQ, *What elicitation techniques are used?*, reveals the elicitation techniques that are used in the projects conducted by the participant subjects. The second question, *What roles are performed in these techniques?*, identifies the stakeholders and organization employees who are involved in elicitation techniques. The third question, *What challenges, if any, are faced in the elicitation process?*, addresses the challenges that the subjects face during the elicitation of requirements. Taken together, all of these questions are directed at the research goal of this study, namely, to identify how requirements are elicited.

*Table 1. Research goal and research questions that are raised in the present study*

| Goal | To identify how requirements are elicited |
|---|---|
| RQ1 | What elicitation techniques are used? |
| RQ2 | What roles are performed in these techniques? |
| RQ3 | What challenges, if any, are faced in the elicitation process? |

***Sampling.*** As noticed by Méndez et al. (2018), there exists great variability in the way that requirements are defined and dealt with, from project to project. Therefore, the aim of this study was to include subjects who are practitioners involved in several software development industrial projects, from different companies.

The target population includes practitioners in charge of RE activities in software development projects. The participating companies were selected from our industrial network. These companies satisfy as many different selection criteria as possible with respect to size, application domain, and business area. In order to obtain their different views regarding their requirements elicitation processes, we initially planned to interview *two* subjects from each company, although at the end we had one company with just one subject and another one with three. Therefore, a total of 24 interviews were conducted to subjects from these 12 companies. Section 4.1 gives further detail on the properties of the interviewed subjects, their companies, and the projects that they were involved in.

***Procedure and Instruments.*** In order to gather data from the target population, we designed a semi-structured interview guide, following the guidelines set forth by Oates (2006). In general, the questions that were included in the guide asked the respondent to focus on a single finished project that s/he was familiar with. The fact that the respondents were asked to focus on a single project, instead of many projects, allowed us to construct a better interpretation and assessment of the contextual information that was available to them. Otherwise, it would have been very difficult to establish any meaningful relationships between the various requirements that engineering practices used and the characteristics of the project for which the requirements were established. The particular project that was discussed by the interviewee was chosen by the interviewee, without any interference from the interviewers. In addition to this narrowing-down of the focus of the interview, we added a number of follow-up questions (such as: *Is this typically how you do this? If not, how do you usually do it?*) in order to identify and understand potentially *representative* practices, as suggested by Lutters and Seaman (2007) and Patton (2002). This approach allowed for a richer vision of the requirements processes undertaken by the interviewees and their opinions to emerge during the interviews. The interview guide is available in Appendix I of the protocol document (see supplementary material). We recorded the content from the face-to-face interviews for later reference. The average duration of each interview was approximately 2 hours in length, of which, approximately 10 minutes were used to describe the company and/or project and 40 minutes were used for the elicitation part (the rest of the time for the other parts of the study, not reported in this paper).

*Data Analysis Procedure.* We applied different coding techniques to analyse the answers (Saldana, 2009) with the support of the Atlas.ti [2] tool. (See details of these techniques in the protocol provided as supplementary material).

We also used statistical techniques to analyse the codes:

- *Contingency tables* to explore frequency data and to perform chi-square tests (Field, 2009).
- *Chi-square test of independence* to test the variety of the sizes of the different contingency tables, as well as more than one type of null- and alternative hypothesis. (The test is considered to be statistically significant if p-value < 0.05.)
- *Cramer's V* (Cohen, 1988) to estimate the strength of the association. Cohen (1988) suggests that, for a large strength of association, the Cramer's V value should be above 0.5. Therefore, when Cramer's V value is greater or equal than 0.1 and smaller than 0.3, it is considered to indicate a *weak* association. When Cramer's V value is greater or equal than 0.3 and smaller than 0.5, it is considered to indicate a *moderate* association. When Cramer's V value is greater or equal than 0.5, it is considered to indicate a *strong* association.
- *Post hoc testing* using adjusted standardized residuals to follow up on our statistically significant results, in order to find which cases are 'responsible' for an association.

In Appendix III, we include the values of *p* and *V* to every statistically-significant and strong association that was found. Remarks on these findings are integrated in the discussion.

## 3.2 Validity

Like all studies in software engineering, the present study faces a number of threats to its reliability. This section outlines these threats in terms of 'construct', 'conclusion', 'internal', and 'external validity', as suggested by (Wohlin et al., 2012). This section then proceeds with a discussion of the corresponding strategies that were used to deal with these threats. Again, we refer to the protocol document (see supplementary material) for a more complete discussion of these issues.

*Construct Validity.* This study is informed by two main principles: (i) rigorous planning and (ii) the establishment of protocols for data collection and data analysis, as suggested by Runeson and Höst (2009). Additionally, the interview guide was piloted before it was used in the field. The pilot interviewees helped us to improve the comprehensibility of the questions, including the use of technical terms that the participants would be familiar with. However, there existed differences in terminology across the different interviews. This issue was addressed by: (i) asking clarification questions during the interviews when needed, and (ii) applying multiple codes to the same statement, so as to capture multiple interpretations. Finally, both in the interview guide and during the actual interviews, the participants were made aware that the information that they provided would be confidential, made anonymous, and aggregated with the rest of interviews. Under such conditions, the participants freely shared their real experiences and perceptions.

*Conclusion Validity.* Throughout the different coding steps, a large number of distinct concepts and their inter-relationships were identified. Traceability from the raw data to the categories was preserved. Different types of triangulation methods were used to minimize possible biases. Different coding techniques ('theory triangulation') were used to capture various aspects of the phenomenon under investigation. Selected cases from the dataset (randomly chosen) were analysed by two of the interviewers so as to identify and to eliminate individual biases when they occasionally arose ('researcher triangulation').

---

[2] http://atlasti.com/

*Internal Validity.* We focused most of the questions on a single software development project. Doing so, we were able to further inquire into, and analyse, specific contexts that had generated a particular elicitation-related decision. To avoid any potential threats to the internal validity of the study, firstly, the interview guide was sent in advance to the respondents. Consequently, the interview participants rarely had difficulty in remembering project details. Secondly, so as to minimize the risk of selecting only successful projects, we informed the participants that the study was not focused on analysing 'incorrect RE practices' but, rather, it was focused on discovering 'how RE is done in industrial practices'.

The fact that the interviews were not transcribed may have represented a threat. However, to mitigate this threat, the audio recordings of the interviews were imported into the qualitative data analysis tool that was used (i.e., Atlas.ti). This offered the same coding functionalities with respect to both the audio files and the text files that were inputted into the program.

To address single researcher bias in the coding process, we applied triangulation in different forms. Selected interviews were analysed independently by two researchers, and the results were then discussed so as to identify and eliminate any individual biases. Responses were triangulated too. In addition, the generated categories were analysed, discussed, and reviewed by the research team so as to ensure the accuracy, understanding, and agreement of such responses.

*External Validity.* Qualitative studies rarely attempt to make universal generalizations that go beyond the studied context. Instead, as Robson explains (Robson, 2002), they are more concerned with characterizing, explaining, and understanding a phenomena as it exists in a particular context. Notwithstanding this caveat, two measures were taken so as to strengthen the external validity of this study. First, we employed a combination of convenience sampling and maximum variation sampling in our selection of the companies (Robson, 2002). Second, we granted the respondents the freedom to choose the project that they wished to talk about during the interview.

## 4 Results

The first subsection below presents the data set that was used in this study. Thereafter, each subsequent subsection presents and discusses one research question that corresponds to the three research questions listed in **Table 3**. Appendix 1 and Appendix 2 provide more complete information about the demographics of the participants and their interview answers, respectively.

### *4.1 Demographics*

In this section, we discuss the most relevant aspects of (i) the subjects who participated in interviews, (ii) their companies, and (iii) the selected projects.

*Subjects.* Most of the 24 subjects had an educational background that is related to computer science, information systems, or software engineering, although a non-negligible proportion of the subjects had graduated in other types of engineering (such as chemical- or civil engineering) or in other areas of science (e.g., telecommunication or robotics). Most of the subjects held either a master's degree or a bachelor's degree. The subjects had between 3 to 25 years of experience in industry (16.2 years on average) and between 0 to 15 years of experience at university or in research laboratory (3.2 years on average). The subjects held different positions of responsibility in their companies, but actively participated in (or were in charge of) RE-related processes (at least for the project they based their answers on). Some of the subjects were new to their position or new to the company (e.g., S11 (F,P11)[3]), while others had significant experience with regards to both their position and the company that they worked for (e.g., S6 (D,P6)).

*Companies.* 12 companies participated in the study. In 11 of these companies, it was possible to interview more than one subject. The software companies included in this study varied in terms of their business areas and size. The 12 companies can be categorized as one of the following: (i) software consultancy companies (SCCs) that perform software development tasks for different clients as their primary business; (ii) IT departments (ITDs) that usually perform or outsource some software development tasks for satisfying the internal demands of an organization; (iii) software houses (SHs) that develop and commercialize specific proprietary solutions.

In addition to the above, some companies explicitly stated that their business area was oriented towards a specific domain. Two of the companies were from the public sector (companies C and L), and the rest of them were private companies.

*Projects.* As explained above, each subject who was interviewed was asked to discuss a single finished project. The resulting set of projects was very diverse in terms of domain, duration, and the number of employees who worked on these projects. The projects were mainly related to embedded systems, websites, mobile applications, and customer business support operations. Regarding duration and size, the various projects took from 4 months to around 10 years to complete, and they involved from 2 individuals up to thousands of people. Only 3 subjects did not know the number of employees who were involved in the project, and one subject made the remark that the number of employees changed along the project's life span. Finally, it is noted that a majority of the chosen projects used a waterfall approach to software development.

### 4.2 RQ1: What elicitation techniques are used?

Many of the interview subjects reported that they used more than one elicitation technique in the project, and these techniques were used in different combinations, e.g. "*Reviewing new platform documentation (what the platform could offer?), reviewing the documentation of the current system (what are we offering right now?), and also end customers' interviews*" (S13(G,P13)).

Consequently, the sum of percentages in this category of question exceeds 100% by far. A summary of the responses is presented in Figure 1.

---

[3] This is the notation that is used in this article to refer to the interview subjects. The notation corresponds to S*x*(*y*,P*z*), where S*x* is a unique identifying number for the interview subject (referring to **Table 1** in Appendix 1), *y* is the identifying number assigned to the interview subject's company (referring to **Table 2** in Appendix 1), and P*z* is the identifying number of the project (referring to **Table 3** in Appendix 1).

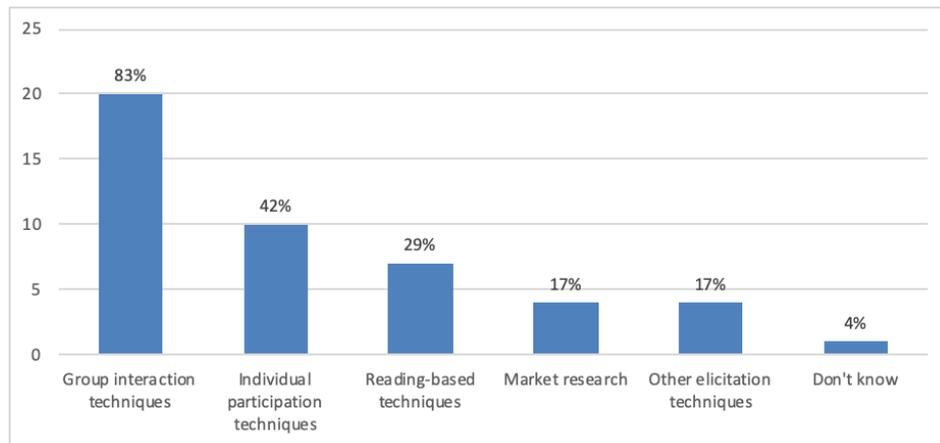

*Figure 1 – Requirements elicitation techniques*

Most of the subjects (83%; 20 respondents) reported that they elicited requirements for their projects by using *Group interaction techniques*. This especially included conducting meetings (14 subjects) and workshops (11 subjects), "*to put everyone on the same page*" (S1(A,P1)). Some organizations based the elicitation process on a rich program of such meetings, as reported by S11(F,P11): "*First, regular meetings at beginning with stakeholders to understand and clarify issues. Second, internal regular meetings of the technical team to specify other requirements. Finally, meetings of the interaction designers with real users that are more related to the user interface*".

However, the interview subjects reported that they also used other elicitation techniques. On the one hand, 10 of them (42%) used *Individual participation techniques*, where only one person is considered at a time. Although, in general, one can think of several potential techniques, the interview subjects only reported on their use of interviews. We thus conclude, for instance, that questionnaires were not used. Interviews were used at different phases of the elicitation process; from the initial phase, to become aware of initial needs, as done by S11(F,P11): "*First, regular interviews at the beginning with stakeholders to understand and clarify issues*", or, later on in the process, to check up on the candidate's set of requirements: "*First, we wrote down a first version of the requirements. Then we met with the business team, by interviewing its members, and tried to check if everything was ok*" (S9(J,P9)). In some cases, these interviews were much more informal in style, for example, consisting of a simple informal conversation or notification, as done by S6(D,P6): "*A customer wants something new; talks to the market unit […] if necessary, the system manager contacts the market unit so it clarifies the doubts with the customer*".

On the other hand, 7 other interview subjects (29%) reported that they used *Reading-based techniques*, which entails the reading of documentation and acting in response to what has been read. Some of the subjects stated that they used these techniques to learn from older, similar systems, as done by S23(L,P23): "*The requirements were at first extracted from the documentation for a project of a similar tunnel in XXX* [anonymous]. *The editorial group was in charge of reviewing the requirements, removing what was not necessary, and adding or modifying requirements. […] Requirements were also gathered by looking at other documentation*. In an extreme case, high-level documentation was not consulted, but rather: "*looking at the code of the old system (because the new system was supposed to be like the old one)*" (S17(I,P17)). Other interview subjects confirmed that they read general documentation about the technologies needed in the project as a reinforcement technique, as illustrated in the following quote: "*Once it is decided the project will be carried out, they specify more detailed requirements, they have meetings inside the team project*

*(system engineers) and dig into the documentation of the machine*" (S14(H,P14)). Only in one case was user feedback considered as an additional input: "*We also looked at the citizens' claims and suggestions documentation, but it was not that much useful*" (S3(B,P3)).

Four of the subjects (17%) mentioned their use of *Market research* as an instrument to gather requirements. S16(I,P16) justified the use of this technique because: "*The project was market driven, so the first high-goal of the system came from market research.*"

Four other interview subjects (17%) outlined a number of additional techniques that they reported using:

- Reuse of requirements: S20(J,P20) reported the reuse of requirements in similar projects of the same domain: "*We have a set of manufacturing requirements ready, that we adapt to every project, … [which are based on] … lessons learned from other projects and market research.*"
- Persona: S3(B,P3) used personas "*to define different specific end customers of the system*".
- Experiments: S19(J,P19) organized "*experiments to get new requirements: things related to new materials, new ways of doing things*".
- External consultancy: S24(L,P24) recounted that the organization "*hired a consultant to extract requirements from this [an initial] document*" due to the complexity of the process.

Finally, 1 subject (4% of the total number of interview subjects) did not know what techniques were used during the elicitation because "*the business requirements are gathered from the customers from another part of the organization*" (S7(D,P7)).

### *4.3 RQ2: What roles are performed in these techniques?*

Figure 2 summarizes the responses given by the interview subjects with regards to the roles that are performed in the execution of RE elicitation techniques. We have created a classification system that distinguishes the roles inside the respondents' organization from roles outside this organization. A number of subsequent subcategories were also created.

We first consider the roles outside the organization. Their involvement is somewhat diverse. Five interview subjects (21%) reported that nobody from outside their organization was involved in requirements elicitation. These subjects' projects were market-driven and the organizations were of the opinion that they possessed the necessary expertise to elicit the new requirements. For instance, the project that was reported on by S14(H,P14) had the goal of decreasing production costs without compromising performance in the development of a machine that formed tetra-bricks. S16(I,P16) selected a project which was had the goal to "*improve a specific part of the safety controller, because a new version of the safety standard they have to follow appeared*". The three subjects who worked for company J (a car manufacturer), i.e. S18(J,P18), S19(J,P19) and S20(J,P20), reported on different ways of working that were always internal to the organization: "*the organization has a set of manufacturing requirements ready, that they adapt to every project*" (S20(J,P20)).

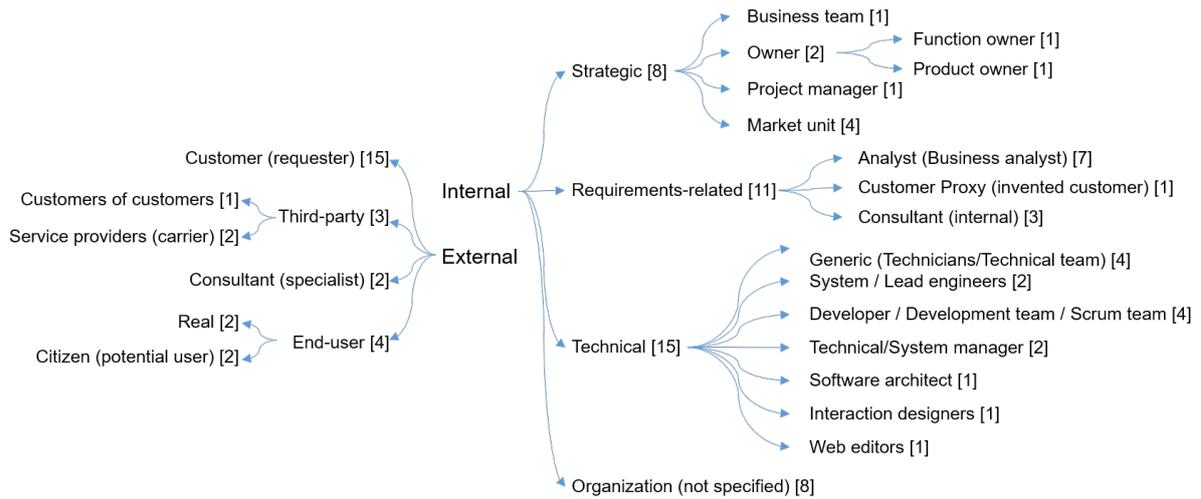

*Figure 2 – Roles that participate in requirements elicitation: A classification*

For the remaining 19 subjects (79% of the total number of interview subjects), the main source of requirements was the customer (also called "client" or "requester" by different subjects), with direct participation reported in 15 cases (62%). In 4 of such cases, the customer's voice was complemented by special roles: customers of the customer (S8(E,P8)), service or carrier providers (S12(G,P12) and S5(C,P5)) or external consultants or specialists, used *"to discuss or clarify how different parts of the project affect to each other"* (S23(L,P23)) and *"to extract requirements from this document [a high-level, general document], because the requirements in that general document were not clearly stated and they were difficult to understand"* (S24(L,P24))[4]. One additional market-driven project did not involve any customer but carrier provider representatives. This was S12(G,P12), who worked for a telecom company on a project that was to offer roaming services to its customers.

Concerning user involvement, only 2 of the interview subjects (8%) stated that real end-users participated directly in the elicitation process, either in the early stages of the process, as with S10(F,P19): *"The requirements were based on market research with the end users"*, or at later stages, as reported by S11(F,P11): *"Finally, meetings of the interaction designers with real users that are more related to the user interface"*. Some of the respondents justified this absence of the end-user because of uncertainty of who the target population would be or because of the size of the target population. Consider S4(B,P4) remarks on this issue: *"[the system] targeted all the Swedish population"*. In 2 other cases (8%), these real users were replaced by potential users, as in the case of S23(L,P23): *"somehow the persons <u>acting</u> as end-customers [were involved in elicitation]"*, and S3(B,P3): *"we looked into citizens' claims/suggestions"*. This last interview subject was involved with more external stakeholders, including the customer and an external consultant.

Regarding the roles that were played in the subjects' organizations, Figure 2 shows greater diversity. We distinguished 3 broad categories of role in the context of a holistic perspective of each organization. This approach is adopted either because some subjects did not know all of the details of the elicitation process, especially in big companies (e.g., S7(D,P7) who indicated that: *"The business requirements are gathered from the customers by another part of the organization"*, or because no clear roles had been defined for conducting the elicitation process. Consider, for example, S4(B,P4)'s observation: *"The workshops involve different people in the organization"*. Quite often, different roles from different role categories collaborated with each other in workshops and meetings, as reported by S2(A,P2): *"meetings between the requester of the system, a developer, the requirements analyst, and the web editors (the ones entering the data on the web"*. However, some organizations preferred to keep certain roles separated, as illustrated by S5(C,P5), who reported: *"workshops with customers (the smartphones makers) and also workshops with the service providers (carriers)"*.

---

[4] Note that both interview subjects work for the same company, a public transport administration.

The role category with the largest involvement in the RE process was the category of technical roles. This included developers (individual developers or a team of developers) but also more specific particular roles, including system engineer, system manager, software architect, interaction designer and web editor. Again, sometimes the subject did not provide enough details on the role specificities in their answers, therefore we considered the "generic" technical person. Usually, individuals with technical roles worked together with others, e.g., in joint workshops as mentioned by S1(A,P1): *"Workshops that involve the development team, the requester and analysts, but no end users"*. The purpose of these workshops was : (ii) to provide their technical perspective: *"internal meetings were held between the requirements engineers at the organization and meetings with technicians to clarify the most important technical points"* (S24(L,P24)); or (ii) to elaborate on the requirements: *"After having this first high-level goal, they [the analysts] had meetings with the developers' team to get implementation proposals on the requirements already defined, getting in that way more technical requirements"* (S16(I,P16)); or (iii) to assess the feasibility of the requirements: *"some architects [from the interview subject's organization] were also participating in the elicitation, so as to be sure the technology fitted the purpose of the system and everything was alright in terms of the technology"* (S13(G,P13)); or (iv) to have a stake on the 'go/no-go' decision: *"the high level goals are passed to the systems engineers, who work out the cost of achieving these high level goals and the benefits this will have. After this study, they decide to carry on with the project or not"* (S14(H,P14)). It is worth mentioning that S1(A,P1) wanted to have separate meetings with only analysts and customers: *"without having the developers' opinion"*.

In terms of individual roles, the role of analyst was most frequently cited by the interview subjects (7 subjects, 29%). This usually referred to a central role, both in terms of conducting the elicitation: *"The business analyst had meetings (interviews) with all stakeholders"* (S12(G,P12)), and in terms of making the transition into the specification phase: *"After that [the elicitation], the requirements analysts write uses cases"* (S13(G,P13)). In the same role category where we placed 'analyst', we included the role of 'customer proxy', as mentioned by S3(B,P3) in his market-driven project where *"the organization did not have direct end customers"*. Consequently, *"a couple of persons in the organization were in charge of deciding what the requirements of the project were going to be"* (S3(B,P3)). As a third role in the same category, 3 subjects (12%) mentioned the participation of 'internal consultants, i.e., people from the organization in charge of the elicitation process who have particular knowledge or skills.

We also noticed that one other role was mentioned frequently, namely, the role of 'market unit' or ('market department'). This role was mentioned by 4 interview subjects (17%). Sometimes, this unit acts a mediator between the customer and the organization: *"a customer wants something new; talks to the market unit; the market unit enters a first high-level requirement in a tool informing which is the system that is* *affected for this requirement; the system manager of this system gets a notification and breaks down the first high level requirement into a requirements specification […] No direct meetings with customers, but by using the market units"* (S6(D,P6)). In other cases, especially in market-driven projects, the market unit had a more proactive role, as reported by S14(H,P14): *"First, there are meetings with the Market Advisor Group (which is the department in charge of defining the future roadmap of the machines). From these meetings the first high level requirements (more like goals) are established"*. The other roles that fall under the same 'strategic' category where we find 'market unit' were mentioned infrequently, including 'business team', 'function owner', 'product owner', and 'product manager'. Both function owners and product owners have particularly significant roles in the requirements elicitation process, note for example, the following remark from S21(K,P21): *"The product owner gathers stakeholders relevant for the project […] and have meetings with them."*

### *4.4   RQ3: What challenges, if any, are faced in the elicitation process?*

As one might expect, most of the interview subjects reported that they faced challenges during the elicitation process on each of the projects that they reported on. In fact, only 2 subjects (8%) declared that they had not been faced with any challenges. They claimed that the lack of challenges was related to having good stakeholders. As articulated by S1(A,P1): *"No, there were no challenges faced because the customer was an educated customer that really knew what she wanted, and understood the limits of the project"*. The remaining 22 interview subjects reported the existence of multiple challenges (with a maximum of 4 challenges experienced by S21(K,P21)). The details of these challenges are summarized in Figure 3 and explained below. Please note that the percentages given in the figure for the 6 categories that were identified correspond to the number of subjects, not the number of challenges. Thus, in the first category, 16 challenges were reported by 13 subjects, i.e. 54%). The figure also shows the average of the relevance of the challenge, according to a 1-5 Likert scale (1: lowest, 5: highest), as reported by the interview subjects.

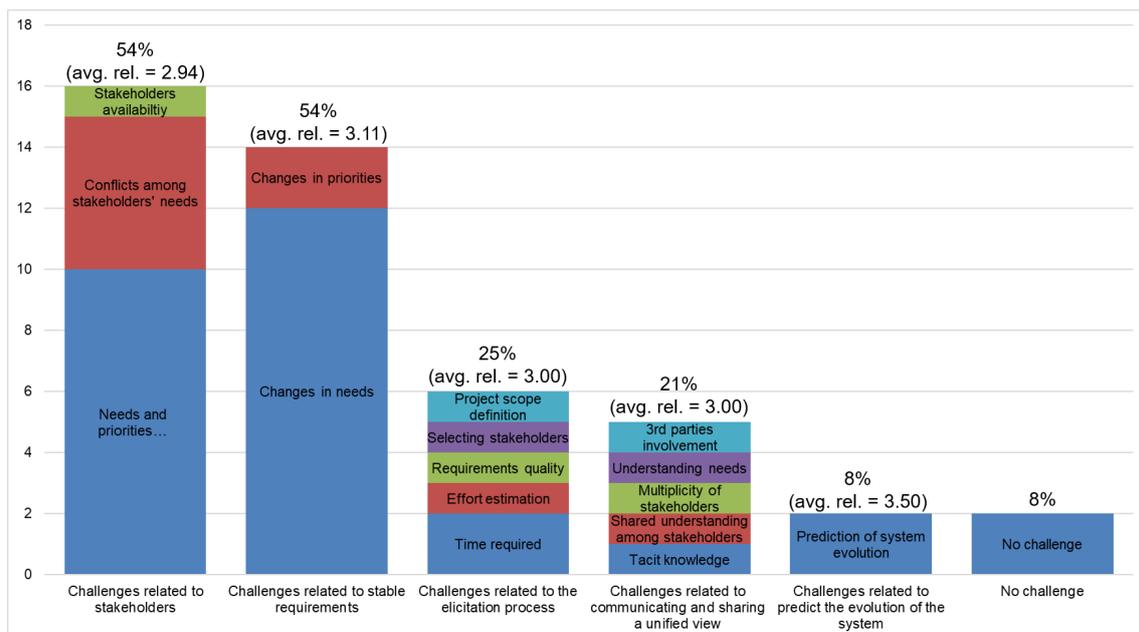

*Figure 3 – Requirements elicitation challenges.*

**(i) *Challenges related to stakeholders*.** To lend support to the claims made above, we note that, the category of challenges that was most frequently encountered by the interview subjects were those *related to stakeholders* (13 subjects, 54%; 2.94 average relevance). In this context, the interview subjects mentioned the following situations:

- The difficulty that stakeholders have in understanding and articulating their needs and priorities was reported as a challenge by 10 interview subjects (42%; average relevance = 3.10). The reasons of this challenge varied vary from project to project. 'Technical reasons' had the lowest impact, as mentioned by S6(D,P6): *"The first challenge is that customers do not know their needs. Needs specially related to the technical aspects, to how the expected behaviour would be affected by other technical aspects"*. In other cases, the reason that is posed is more closely related to the stakeholders possessing poor knowledge about the initial state: *"The main problem was that the stakeholders did not know their needs. As it was a migration, the system was expected to do the same as before (same functionalities) but built over the new platform [...]. Nobody knew how the old system was working and the customer did not know what changes they wanted exactly (the things to improve from the old version) until they saw at the tests of the new system that things didn't work as expected"* (S17(I,P17)). Such a situation may have serious consequences, *"as it implies more costs, delays, etc."*. As a consequence of this, its relevance was considered to be greater ( S17(I,P17) rated the relevance of this reason with a '4').

- The emergence of conflicts among stakeholders was reported on by 5 interview subjects (21%). The reasons why such conflicts took place were diverse. For instance, S5(C,P5) reported that: *"what the customer wanted was not possible to be implemented because of the restrictions imposed by the smartphone or the service provider's technical requirements"*. Sometimes, conflicts arose as a consequence of lack of knowledge, as stated by S19(J, P19): *"Stakeholders did not know their needs and conflicts among needs happened, because there were too many options available for accomplishing a goal, and at some moment it was necessary to decide and stop getting new options"*. The relevance of this challenge was not considered to be particularly high (average relevance = 2.80). One reason why this challenge was not considered to be of high relevance may be associated with the fact that such challenges can be quickly resolved: *"at the end the person with money decided and the differences where solved mostly at the early stages of the project"* (S12(G,P12)). Exceptions to this were voiced by S18(J,P18) and S19(J,P19). The fact remains, however, that these two interview subjects ranked all challenges with a '4', thereby suggesting they faced problems in general with respect to the elicitation process with challenges retrofitting each other.

- The importance of having customers who are *available* to participate in the elicitation process. This challenge was only reported on by (S9(E,P9)): *"The workload of the stakeholders is a challenge, because they were so busy that they were less available to get some details needed to get the requirements"*. This challenge was successfully dealt with in this project, as suggested by (S9(E,P9))'s relevance score of '2': *"The impact was low, because the customer management saw the project as really important and they lift off work from the workers, so they were available to answers questions"* (S9(E,P9)).

The main consequence of stakeholders not knowing their own needs results in more time being spent on the elicitation process than expected. This scenario is described by S8(E,P8) in the following: *"We faced the problem of stakeholders not knowing what their exact needs were, and we spent too much time in the elicitation of the requirements because of that, because it is difficult to get requirements from the customers. They don't know how to separate what is important and what is not, so it is difficult to get the right requirements from customers from the beginning"*.

**(ii) Challenges related to stable requirements.** 13 interview subjects (54%) stated that the *stability of the requirements* and, more specifically, *a change of needs* (12 subjects, 50%) and *a change in priorities* (2 subjects, 8%), posed challenges to the elicitation process. The average relevance was a bit higher than above (3.10). A maximum score of '5' was recorded by S6(D,P6): *"The impact was very high, and at the end a new project had to be developed"*. S22(K,P22) scored this type of challenge at '4': *"Especially if the changes / missing features had a big impact on other things on the system, and some extra functionalities were moved to the maintenance stage to be able to finish on time"*. In contrast, this challenge was of less relevance to S15(H,P15) who scored it at '2': *"The impact was rather low, just small changes or additions were done"*. Most of the time, changes were related to stakeholders who had changed their mind during elicitation process, but other reasons exist why changes were made, as illustrated by S18(J,P18): *"Instead of being problems between the stakeholders, it was more about changes in the market or solving problems found during the integration of the different software or modules, as the requirements were being specified in a 3-year period, and, technologically, things change really fast"*. The consequences of such changes are mainly increased cost and delays, although one interview subject, S19(J,P19), reported on: *"less quality of the requirements (because they did not delay the project, but used less time for some of the tasks)"*.

**(iii) Challenges related to evolution of the system** are also related to the dimension of time, but from a different perspective. 2 interview subjects (8%) mentioned that it was difficult to *predict the evolution* of the system. This challenge may have arisen from wanting to have a system that would last a long time, as S9(E,P9) explained: *"As we were deciding requirements for a system that would be working on the market for 10-15 years, it was really hard to get an answer on how the business will look in the future. Will market changes affect the system (as it will have a large life cycle)? In that sense, although there was a business plan and everything, it was difficult to establish"*.

In other cases, this challenge can be caused by the appearance of new technologies during the project, as was in the case reported on by S23(L,P23): *"The big challenge was to know if requirements will be good in some years, especially in terms of technical systems, because new technologies might emerge during the requirements elicitation that are better for the project"*. For these two subjects, the relevance of this challenge was quite high, '4' and '3' respectively. These interview subjects scored this challenge higher than the categories above.

**(iv) Challenges related to the viewpoint.** 5 interview subjects (21%) claimed that *communicating and sharing a unified view* of the project was also challenging (average relevance = 3.0). S7(D,P7) scored this challenge with a '4', thus indicating that it can be experienced as quite a severe challenge: *"The main problem *is understanding the main goals (the business requirements) that the project is in charge of. The impact is high, since they could end developing something not right*".

In a less critical form, the challenge can be posed as S10(F, P10) explained: "*The first challenge was putting customers (the marketing department) and the requirement engineers on the same page according to the project, that is, arriving to a common view and a common vocabulary so they understand each other properly*".

The subjects reported misunderstandings in other contexts:

- The multiplicity of stakeholders: "*The last challenge was having the difficulty to communicate with different departments at the stakeholder site, to check that everything was properly specified (for instance, with the specifications of the prices, and how it was worked out by the system and so on*" (S9(E,P9)).

- The involvement of a third party company which is not always available for consultation in the elicitation process: "*The first challenge is the communication with the 3rd party for the API; this 3rd party is designing an API for the same system to be used for mobile applications, but the intermediary is the customer organization, so they do not communicate directly with the 3rd party for API design*" (S13(G,P13)).

- The existence of tacit knowledge: "*[Stakeholders] took it for granted that some aspects would be developed, but as they did not pass this knowledge to the organization, some requirements were missing at the end*" (S22(K,P22)).

*(v) Challenges related to the process.* Finally, 6 interview subjects (25%) admitted to the existence of challenges that are *related to the elicitation process* per se (average relevance = 3.0). The following explanations were provided:

- "*Defining the scope of the project*" (S3(B,P3)), did not score as highly relevant (relevance = '2'), because "*once they decided what to develop, they continued with that*".

- "*Project estimation was not good*" (S8(E,P8)), was considered to be a severe problem (relevance = '4'), because it was necessary for the interview subjects "*to cut some functionalities and the budget of the project increased considerably*".

- "*Too much time [with respect to initial schedule] spent in elicitation*" (S13(G,P13)), was considered to be of a low relevance (relevance = '2'), because "*the time scheduled for elicitation was too short […] it was known from the start*".

- "*The main challenge was that the quality of the initial requirements was not good enough*" S20(J, P20). This challenge was scored as 'fairly' relevant, (relevance = '3') because "*there existed the possibility that relevant requirements were missing*".

- "*Selecting the right stakeholders was a challenge*" (S21(K,P21)). This challenge was subject to varying relevance scores, from '2' ("*only clarifications were needed*"), and '4' ("*more cost and delays*").

# 5 Analysis

The three subsections below address one research question each. We again refer the reader to the three appendixes at the end of this paper to learn more about the analysis that is reported here.

## 5.1 RQ1: What elicitation techniques are used?

As commented on above, *group interaction techniques* was the most frequently used approach for RE by our respondents. This observation is in agreement with Méndez et al.'s study (2015), where the most commonly used elicitation technique was reported to be *the use of workshops*. This is also very similar to Wagner et al.' study (2017) where *meetings* were the second most frequently mentioned technique. However, a fundamental difference between the present study and these two studies exists. In the present study, our interview subjects reported on *a lack of stakeholder involvement*, which was not the case in Méndez et al. (2015) and Wagner et al.'s (2017) studies. In these studies, stakeholder involvement was reported as commonplace. We did not find any statistically-significant explanation for this difference. Therefore, to identify the root cause of this difference, we paid attention to two determinants that may influence the way in which requirements are elicited: the 'company's business area/area of operations' and 'the methodologies that the company uses'. In our study, half of the companies (A, B, D, G, I, and L) are mainly 'market-driven' companies because of the business area they focus on. This may imply a lower degree of stakeholder involvement in the requirements elicitation process (actually, 8 interview subjects from these companies reported that they did not involve their customers in their elicitation process. Unfortunately, neither Méndez et al. (2015) nor Wagner et al. (2017) mention the business area of the companies that they studied. Concerning methodologies, we find no difference in relation to the RE techniques that are used by companies which use 'agile' methodologies versus companies which use more traditional operational methodologies.

Other empirical studies have analyzed the elicitation techniques that are used by different companies (see Table 4). In the study conducted by Todoran et al. (2013), *interviews*, *questionnaires*, *analysis of existing documentation*, *surveys*, and *prototyping* were the most frequently applied requirements elicitation techniques. In Todoran et al. (2013), *group interaction techniques* was not the most common approach (as it is in this study) these researchers focused on cloud environments, where stakeholders are distributed, thus making difficult to gather stakeholders together and conduct the requirements elicitation process.

It is worth mentioning that, in spite of the dominance of *group interaction techniques* in our study, only 2 out of the 20 interview subjects used them in isolation. The rest of interview subjects used *group interaction techniques* together with *interviews* (6 subjects), *reading-based techniques* (5), *market research* (4), and *other techniques* (3).

Although not reporting on actual industrial practices, Pacheco et al.'s (2018) secondary study provides interesting insights into requirements elicitation techniques, as reported on in the scientific literature. They reviewed a total of 140 studies, of which 92% included some type of empirical assessment (case studies, experiments). From the 140 studies, 109 of them were used to answer the research question: *What mature techniques are used to elicit requirements?* The results of this meta-study provide evidence that the most common techniques in software requirements elicitation are: *traditional techniques* (corresponding to *interviews* and *surveys*; 38 out of 109 studies, 36%) and *collaborative techniques* (almost equally distributed among *focus groups*, *workshops*, and *brainstorming*; 13 out of 109 studies, 12%). Our study achieved similar results. The most frequently used approach was *interaction techniques* (83%) (which is equivalent to *collaborative techniques*) and, in the second position, in terms of frequency, *individual participation techniques* (38%) (which is equivalent to *traditional techniques*). The difference in the relative ordering of these two approaches might be due to the context of the different research efforts; for instance, differences in the domain of the projects or the type of organization. We note, however, that these different contexts are not reported on in sufficient detail in Pacheco et al.'s study. This is probably due to the high number of primary studies they consulted. Pacheco et al. (2018) do highlight the fact that, in 24 of the 109 studies (22%), more than one requirements elicitation technique was used. This observation prompts them to recommend the use of more than one requirements elicitation technique. This is corroborated by our study, were the majority of interview subjects (20 subjects, 83%) use more than one elicitation technique.

*Table 1 – The most common elicitation techniques identified in our study and in other empirical studies*

| | | | *Industrial empirical studies dealing with elicitation techniques* | | | |
|---|---|---|---|---|---|---|
| | | | *(Méndez et al. 2015)* | *(Todoran et al. 2013)* | *(Wagner et al. 2017)* [5] | *Our study* |
| *Elicitation techniques* | *Group interaction techniques* | *Workshops* | 80% | - | - | 46% |
| | | *Meetings* | - | - | ≈71% | 58% |
| | *Questionnaires* | | - | 52% | - | - |
| | *Documentation analysis* | | - | 74% | - | - |
| | *Surveys* | | - | 47% | - | - |
| | *Prototyping* | | 44% | 68% | ≈74% | - |
| | *Scenarios* | | - | - | - | - |
| | *Interviews* | | - | 79% | ≈76% | 42% |
| | *Change requests* | | 58% | - | - | - |
| | *Stakeholder involvement* | | HIGH | LOW | HIGH | NOT ALWAYS |

In addition to the above, we found a number of significant correlations (see Appendix III for p-values and Cramer's V values)

- Some job positions were more likely to use a particular technique. Whilst project managers and software architects used *group interaction techniques*, consultants used *individual participation techniques* (*interviews*, mainly). Product/function owners, on the other hand, used *market research*. Given the small number of interview subjects who were in these job positions, the robustness of these correlations is moderate. However, we wish to claim that they do suggest that the role that an engineer plays in the organization will influence the choice as to which techniques are chosen.
- All four respondents who used *market research* for elicitation mentioned *instability of requirements* as a challenge. The reason behind this unanimous agreement on this point may be related to the rapid obsolescent of *market research* when requirements evolve, thereby demanding that an update of the analysis be done, so as to accommodate such changes.
- The two strongest correlations were found with respect to certain project characteristics. On the one hand, *the number of employees in the project* is strongly correlated with the use of *individual participation techniques* -- such techniques are preferred in small projects (n < 10). One hypothesis regarding this observation is that small projects create an atmosphere that is conducive for individuals to elicit requirements. On the other hand, the use of different elicitation techniques was mentioned by interview subjects who were involved in by projects with large budgets (> 1B euro). Enjoying such a large budget allows the company to put in place different actions as to conduct experiments or to hire external consultancy.

---
[5] Wagner et al. (2017) do not provide figures in their results and, consequently, these figures had to be extracted from the graph provided in their paper.

### 5.2   RQ2: What roles are performed in these techniques?

The previous research that is mentioned in Section 2 has not given any significant attention to the roles that are involved in performing elicitation. The only aspect that touches on this (and is mentioned in some studies) is the collaboration that takes place between internal- and external stakeholders. For instance, Sethia and Pillai (2014) propose two dimensions that are related to the meetings that take place between project teams and users, while Todoran et al. (2013) mentions the different ways brainstorming sessions can be performed, depending on whether they involve only the project team or the customer too. Other studies mention roles at the same general level, but omit to focus on individual roles, as we do in the present study. Given this difficulty with respect to comparing existing work in this area, we explore possible correlations among roles and all the other variables in our study. By doing this, we discovered several statistically-significant findings that are of interest:

- All of the 8 subjects with a requirements-related job position (i.e., 'business analyst' or respondents with either 'requirements' or 'consultant' in their job description) involved the customer in the elicitation process. One explanation for why this is so because experienced requirement engineers (and similar positions) are aware of the severe consequences of not having the customer 'in the loop', as it were, while individuals who play other roles in the organisation, and especially technical roles, may not be equally aware of this need.
- The only two interview subjects who reported the involvement of external consultants had also worked on projects in the construction domain that had costs higher than 1B €. In other words, hiring specialists for the elicitation process seems not to be a priority for these two subjects.
- The only two subjects who reported the involvement of real end-users followed an 'agile' methodology in their projects. Although one might have expected the involvement of end-users in projects following 'waterfall' methodologies (since they usually have a longer requirements elicitation stage), that was not the case in the projects two subjects reported on.
- There exists a correlation between the roles involved in the requirements elicitation process and the number of years the subjects had been working for in their organizations or positions. In particular, we note that all 4 subjects who had generic technical roles had been working for their organization for less than 5 years (and therefore they had been working in their specific job position less than 5 years, too), whilst half of the people who were involved in their organization's elicitation process had been working for their organization for more than 15 years. Both relations may be related to the level of knowledge of the people inside the organization who are approached for such a task. In the first case, the relatively less-experienced interview subjects reported that they approached the technical people (a role) in a specific way, because of their lack of knowledge of how their organization works from the inside. In the second case, the relatively more experienced interview subjects approached other roles inside the organization in a general way, because, after working for so many years in the organization, they knew which specific individuals they should approach.

- The only two interview subjects who reported on using external consultants also used a (business) analyst in the elicitation process. This significant correlation indicates that organizations feel it necessary to appoint business specialists to interact with consultants, probably because they can speak to each other from a similar perspective.
- All of the four interview subjects who stated that they used market units during the elicitation of requirements claimed that they faced a number of challenges related to stable requirements. This might be due to the fact that market units change the requirements quite often (based on direct customer petitions or the results of a market analysis), and therefore this affects the stability of requirements. In contrast, almost all of the subjects who reported on involving the organization in the elicitation process (6 out of 8 people) were not faced with similar challenges related to stable requirements.
- In relation to the interview subjects who stated that they had worked in market-driven projects which did not involve end-users in the process, we claim that such individuals might not be seen as 'requirement engineers' *per se* but, instead, as product managers or product owners performing much of the RE work needed in their companies.

### 5.3 RQ3: What challenges, if any, are faced in the elicitation process?

A further analysis was conducted where the personal characteristics of the subjects were taken into consideration. In this regard, our findings were:

- Challenges related to predicting the evolution of requirements elicitation are faced by subjects who only hold a bachelor's degree (2 people). The reason for this may be associated to a lack of advanced knowledge related to 'requirements', which is usually only acquired in specialization studies, such as certain MSc degree programs.
- Almost all of the interview subjects who reported that they faced challenges related to the elicitation process (5 out of 6 people) had been working in the same position for more than 10 years. These results are somewhat surprising. One might expect that higher levels of experience would result in better elicitation management and either the avoidance of problems of the speedy resolution of such problems. Possible causes behind this association could be:
    - In a rapidly evolving field like requirements elicitation, we note that, in the context of data-driven software engineering, the knowledge that professionals possess can quickly become outdated if they do not make an effort to master new techniques (Allen and Grip, 2007). A person with a great deal of experience in a certain field, especially if they are older, may be more reluctant to acquire such new skills (Maurer, 2001).
    - Performing the same type of activity may become repetitive and lack excitement for the individual, thereby impacting negatively on the person's levels of motivation and, in turn, on the work done (Crowley, 2011).
- Almost all of the interview subjects who said that they faced challenges that were related to communicating and sharing a unified view (4 out of 5 subjects). We note that their projects' domain was carrier systems (either a business system or an internal system). A possible explanation for this challenge could be a lack of common, shared terminology (i.e., a standard glossary) for this domain, thus making difficult for these professionals to come to a common understanding of the project on hand and further interfering their understanding of the requirements. In this context, it is worth highlighting the fact that this domain is very technical, so if all of the people who are involved with the requirements do not have the same level of knowledge (including the stakeholders who are involved in the process), this might hinder understanding of the requirements.

- Challenges related to stakeholders are more likely to appear in projects that follow a 'waterfall' methodology.11 out of the 13 subjects who faced challenges use this methodology in their projects. This might be due to several factors:
    - 'Agile' methodologies use techniques, such as storytelling and poker planning, that are able to cope with the two main problems related to stakeholders: i.e., their difficulty in articulating their needs and the conflicts among the stakeholders.
    - Requirements elicitation is a process that proceeds until the end of the projects, whilst, in 'waterfall' methodologies, requirements elicitation is typically done at the beginning of the project only. Consequently, in 'agile' methodologies, stakeholders are involved until the end of the project, in contrast to 'waterfall' methodologies, where stakeholders are involved just at the beginning of the project. The continuous involvement of stakeholders in 'agile' methodologies may be a factor in reducing the challenges related to stakeholders.
    - 'Agile' companies apply techniques and strategies, such as A/B testing, and frequent product releases that can be considered as a form of data-driven elicitation approach.

Taken together, all of these factors may explain the fact that challenges related to stakeholders are more prone in projects that use 'waterfall' methodologies than they are in projects using 'agile' methodologies.

Challenges related to communicating and sharing a unified view are more likely to be faced by the individuals involved in projects that do not include the business team in the elicitation process. 4 out 5 interview subjects who are faced with this challenge did not involve the business team. This may be explained by the fact that business team members are also educated stakeholders who have a good understanding of the domain of the system and may act as facilitators for the rest of the stakeholders. Our results are corroborated by some of the studies reported in Section 2 (see Table 5). Four of these studies (Hiisilä et al., 2015; Méndez et al., 2015; Méndez et al., 2016; Sethia et al., 2014) identify challenges that are also related to project stakeholders. Challenges related to the communication and sharing of a unified view can also be detected in Bjarnason et al., 2011; Hiisilä et al., 2015; Méndez et al., 2016; Raatikainen et al., 2011 and Sethia et al., 2014. Moreover, Hadar et al. (2014) claim that the level of domain knowledge affects customer communication and a proper understanding of their needs during requirements elicitation via interviews.

Related to this communication challenge, Bjarnason et al. (2011) demonstrate that one of the causes why communication gaps exist is because individuals may have an unclear vision of the overall goal of the project. This is also identified as a challenge in our study. These 'challenges of scope' (which, in the present study, are part of the challenges related to the elicitation process) are also identified by Hiisilä et al. (2015). Bjarnason et al. (2011) state that communication gaps usually imply that there exist low levels of motivation to contribute to requirements work. In this study, this observation falls under 'challenges related to stakeholders'. Hiisilä et al. (2015) also identifies challenges that are related to 'co-operation of the stakeholders', in RE.

To conclude, we note that Méndez et al. (2015, 2016) and Sethia et al. (2014) support our findings about the existence of challenges related to stable requirements, while Bjarnason et al. (2011), Méndez at al. (2015), and Méndez et al. (2016) similarly do so with respect to the existence of challenges that are related to the quality of requirements. (Méndez et al. 2016) also identifies the time that is required to perform the elicitation work as a challenge. (Both of these latter challenges fall under the category 'challenges related to the elicitation process' in our study).

It is worth remarking, however, that none of the studies mentioned here have identified 'tacit knowledge' (covered in our study under the category 'challenges related to elicitation').

*Table 2. Elicitation challenges identified in the present study which are supported by other empirical studies*

| | | *Empirical studies dealing with elicitation challenges* | | | | | | |
|---|---|---|---|---|---|---|---|---|
| | | *(Bjarnason et al. 2011)* | *(Hadar et al. 2014)* | *(Hiisilä et al. 2015)* | *(Méndez et al. 2015)* | *(Méndez et al. 2016)* | *(Raatikainen et al. 2011)* | *(Sethia et al. 2014)* | *Present study* |
| *Elicitation challenges* | *Challenges related to stakeholders* | - | - | $X^5$ | $X^{5,6}$ | 25% (avg. rel. = 2.92)[7] | - | $X^5$ (avg. rel. = 3.07)[7] | 54% |
| | *Challenges related to the stability of requirements* | - | - | - | $X^{5,6}$ | 33% (avg. rel. = 3.08)[7] | - | $X^5$ (avg. rel. = 3.15)[7] | 54% |
| | *Challenges related to the elicitation process* | 33% | - | $X^5$ | $X^{5,6}$ | 48% (avg. rel. = 2.86)[7] | - | $X^5$ (avg. rel. = 3.15)[7] | 25% |
| | *Challenges related to communicating and sharing a unified view* | 100% | $X^6$ | $X^5$ | $X^{5,7}$ | 41% (avg. rel. = 2.71)[8] | $X^5$ | $X^5$ (avg. rel. = 3.25)[7] | 21% |
| | *Challenges related to predicting the evolution of the system* | - | - | - | - | - | - | - | 8% |

## 5.4   Other aspects (tools, processes)

During our analysis of the tools used during the RE process in the projects reported on by the interview subjects, we note that none of the interview subjects reported on using a particular tool for performing the elicitation *per se*. Given that companies have a natural tendency to investigate whether certain tools can assist them when they face challenges, we conclude that the current market needs better tools to support requirements elicitation. This need may be considered all the more pressing when we note that all of the companies that were involved in this study follow well-established practices. In fact, most current RE tools are not actually focused on the elicitation of requirements, but on their specification and management. Other tools that are commonly used during RE (but which are not specific to RE, e.g., Microsoft Word and Microsoft Excel) do not really give support for the elicitation process itself, which again, lends support to our argument.

---

[6] No figures on the number of participants who reported these challenges are provided in the original studies.

[7] Only the mode, median, and level of statistical significance is provided. Mode and median were bigger than 3.5 on a Likertscale from *1- Totally disagree* to 5 – *Completely agree*. Statistical significance is reported in terms of p-values.

[8] '*avg. rel.*' refers to the average of the relevance of the problem stated by the participants. The closer the value is to 1, the more important it is.

Remarkably, given the preference of our respondents for group interaction techniques, tools for implementing collaboration and supporting such collaborative workshops could be of use to requirement engineers. However, none of the interviewees acknowledged the existence of such tools, which can be considered to be another barrier to be overcome during requirements elicitation.

We endeavoured to gather evidence in our interviews that was related to the possible adoption of tools in the future. We received only a few hints along this line of thought, but the most remarkable piece of information we received was that some interview subjects would like to have an overview of current existing solutions inside the company (so that the analyst can have a clear overview of other systems that might be similar to the one currently in use or to inspire this individual in the elicitation of requirements). For instance, S1(A,P1) declared that *"We are looking now at the possibility of creating some kind of repository with this tool [Power Designer] to have all the knowledge about all the projects centralized, so we know what each system is about"*. In a similar vein, S12(G,P12) stated that *"Right now the tool [Power Designer] is only used for managing the requirements at a project level, although it is expected to be used in the future to provide an overview of all the existing solutions inside the organization"*.

## 6 Conclusions and Future Work

In this paper, we have presented the findings of an interview-based, empirical study that was conducted at 12 Swedish companies, involving 24 practitioners. The study has answered three research questions:

- Except in small projects, group interaction techniques (meetings and workshops) were the RE techniques that were preferred over the rest, (including individual participation techniques – interviews and questionnaires–, reading-based techniques, and market research). Although we note that these techniques were rarely used alone.
- Customers and technical people (developers, system architects) constituted the roles which evoked the highest levels of participation in the RE process from outside and inside the organization, respectively. Market-driven organizations usually did not involve stakeholders from outside the organization, and user involvement was, in general, limited.
- Challenges related to stakeholders and requirement stability (especially in large organizations) were dominant over other type of challenges (such as challenges related to the elicitation process, communication issues, and challenges with adopting a long-term view of the system).

Some of the above findings align well with previous studies, but others are somehow conflicting. We have speculated about possible reasons why these discrepancies are present (for example, we observe that market-driven projects are more numerous in our study than in others). It is also worth mentioning two particular areas for which our study provides details but which are not found in others. They are (i) a detailed study of the participation of different *roles* in the elicitation process and (ii) a two-level classification of challenges found in requirements elicitation.

As future work, we plan to complement the results reported on in this paper with those obtained in relation to 'requirements specification', which was also analyzed as part of the same study. Any findings that relate elicitation techniques, roles, or challenges with specification approaches will shed additional light to the complex scenario of requirements elicitation in industry. Another line of research that we are particularly interested in following is to be found in the evolution of the field in the emerging domain of the Internet of Things and self-adaptive systems, where the availability of enormous amounts of data drives practitioners to adopt data-driven approaches in combination with the traditional techniques and roles that are mentioned in this paper.

Finally, the study reveals to us the fact that the large body of RE elicitation research that now exists has yet to inform industry practices. The limited use of research results may be due to problems with the technology transfer models that are in place within the industry (for instance, practitioners may simply be unaware of the existence of modern elicitation techniques), but it may also be the case that the methods that have been developed are not usable and/or useful for practitioners (given, for example, problems related to scalability, or the necessity of training people in the field). Consequently, we argue that future work in the area of RE should also address the gap between research and practice.

# Appendix I – Description of population

*Table 1 – Subjects included in the empirical study*

| ID | Highest Level of Educational Attainment | Years in Industry | Years in University or Research Labs | Job Position | Years in Position | Years in Organization |
|---|---|---|---|---|---|---|
| S1 | BSc in Computer Science | 15 | 3 | Business Analyst | 3 | 3 |
| S2 | MSc in Computer Science | 15 | 3 | Project Manager | ≈5 | 10 |
| S3 | BSc in Information Systems | 20 | ≈4 | System Analyst | 6 | ≈9 |
| S4 | BSc in Computer Science | 13 | 3 | Requirement Analyst | 13 | 13 |
| S5 | MSc in Computer Science | 25 | 5 | Requirement Analyst | 2.5 | 4 |
| S6 | BSc in Information Systems | 20 | 0 | System Manager | 15 | 20 |
| S7 | MSc in Computer Science | 19 | 5 | System Manager | 6 | 19 |
| S8 | BSc in Computer Science | 15 | 0 | Senior Project Manager | 15 | 15 |
| S9 | BSc in Energy Systems | 20 | 0 | Senior Business Consultant | 6 | 6 |
| S10 | MSc in Computer Science | 16 | 0 | Senior Developer | 9 | 9 |
| S11 | MSc in Software Engineering | 17 | 5 | Unit Manager | 0 | 0 |
| S12 | MSc in Business | 12 | ≈5 | Solution Designer | ≈8 | ≈9 |
| S13 | BSc in Computer Science | 23 | 0 | Business Analyst | 14 | 14 |
| S14 | PhD in Food Engineering | 10 | 15 | System Engineer | 2 | 5 |
| S15 | MSc in Chemical Engineering | 10 | 0 | System Engineer | 0.25 | 4.5 |
| S16 | BSc in Telecommunication | 25 | 0 | Product Manager | 5 | 19 |
| S17 | MSc in Industrial Engineering | 8 | 0 | System Engineer | 8 | 8 |
| S18 | MSc in Computer Vision and Robotics | 9 | 5 | Project Leader | 2 | 2 |
| S19 | MSc in Electrochemistry and Electronic Sensors | 3 | 3 | Lead Engineer | 0.5 | 2 |
| S20 | PhD in Civil Engineering | 23 | 10 | Software, Manufacturing & Electrical Engineer | 1.5 | 16 |
| S21 | MSc in Computer Science | 21 | 0 | Senior Consultant | 5 | 12 |
| S22 | BSc in Interaction Design | 9 | 3 | Senior Consultant | 3 | 9 |
| S23 | BSc in Quality Engineering | 15 | 4 | Assignment Manager | 5 | 4.5 |
| S24 | BSc in Mathematics, Physics and Computer Science | 26 | 4 | Requirements Engineer | 3.5 | 3.5 |

*Table 2 – Companies included in the empirical study*

| ID Organization | ID Respondent | Number of Employees | Main Business Area |
|---|---|---|---|
| A | S1, S2 | ≈2,000 WW | ITD of a Telecommunication Operator |
| B | S3, S4 | ≈900 | SCC in the Public Sector |
| C | S5 | ≈350 | SH (UI Platforms for Symbian-Based Smartphones) |
| D | S6, S7 | ≈115,000 WW | SH (Telecommunications Products) |
| E | S8, S9 | ≈68,000 WW | SCC |
| F | S10, S11 | 50 | SCC |
| G | S12, S13 | 800 | SCC (Telecommunication Products) |
| H | S14, S15 | ≈23,000 WW | ITD of a Tetra Bricks Manufacturer |
| I | S16, S17 | ≈150,000 WW | SH (Power and Automatization Systems) |
| J | S18, S19, S20 | ≈20,000 | ITD of a Car Manufacturer |
| K | S21, S22 | 1,200 | SCC |
| L | S23, S24 | Not sure | Public Transport Administration |

*Table 3 – Projects included in the empirical study*

| ID Project | ID Subject | Project Main Functionality | Project Domain | Project Duration (in years) | Project Number Employees | Project Costs (in €) | Project Methodology |
|---|---|---|---|---|---|---|---|
| P1 | S1 | Getting customer feedback | Messaging System | 1 | ≈10 | Not sure | Waterfall |
| P2 | S2 | Webshop for acquiring phones and contracts with a carrier | Website | 1 | ≈10 | Not sure | Waterfall |
| P3 | S3 | Translating a website into English | Website | 1 | 10-12 | ≈0.5 Millions | Agile |
| *P4* | S4 | Management of the social security rights of children | Website | 1.5 | ≈35 | ≈6.0 Millions | Agile |
| *P5* | S5 | OS for a specific smartphone taking into account the carrier's restrictions | Mobile OS | 0.5 | ≈100 | Not sure | Waterfall |
| *P6* | S6 | Carrier system to track the users' consumption | Machine to Machine System | 0.25 | 7 | Not sure | Agile |
| *P7* | S7 | Providing services to customers (charging, changing plan, consumption, etc.) | Carrier Business Support System | 2.5 | Not sure | Not sure | Agile |
| *P8* | S8 | Managing consumption energy levels measured by energy companies | Energy Measurement System | 1.5 | ≈2 | Not sure | Waterfall |
| *P9* | S9 | System for an energy company involving the contract and offering module | Business Support System | 2 | Not sure | ≈1.3 Millions | Waterfall |
| *P10* | S10 | System for a carrier involving big data, call data management, contracts management, etc. | Carrier Internal System | 1 | 100 | Not sure | Agile |
| *P11* | S11 | Webshop for acquiring public transport system tickets | Website | 0.33 | 5 | Not sure | Agile |
| *P12* | S12 | Offering roaming services to customers | Carrier Business Support System | ≈1.5 | ≈20 | ≈1.5 Millions | Waterfall |
| *P13* | S13 | Managing customer calls in a customer service centre | Carrier Internal System | 1.5 | 25 | Not sure | Waterfall |
| *P14* | S14 | Modifying an existing machine (and its internal system) to make it more productive | Embedded System | 4 | 35 | ≈9.0 Millions | Waterfall |
| *P15* | S15 | New machine (and internal system) for a new package | Embedded System | 0.75 | 10 | Not sure | Waterfall |
| *P16* | S16 | Managing control and safety processes | Embedded System | ≈1.5 | ≈200 | ≈6.0 Millions | Agile |
| *P17* | S17 | Controlling the machines of a sugar factory | Embedded System | 1.5 | 6 | Not sure | Waterfall |
| *P18* | S18 | Managing the different functionalities of a car | Embedded System | ≈3 | ≈60 | Order of Billions | Waterfall |
| *P19* | S19 | Controlling the charge of the battery in electric cars | Embedded System | 2 | 20 | Not sure | Waterfall |
| *P20* | S20 | Controlling the machines for producing a car | Embedded System | 6-7 | x000 | Order of Billions | Waterfall |
| *P21* | S21 | Checking films, book tickets, etc. for a cinema company | Mobile App | 1 | 18 | ≈1.5 Millions | Waterfall |
| *P22* | S22 | Integrating payment services | Mobile App, Website | 0.25 | 12 | Not sure | Agile |
| *P23* | S23 | Specifying a tunnel construction details and safety systems | Construction | 10 | Not sure | ≈2.5 Billions | Waterfall |
| *P24* | S24 | Specifying a tunnel construction details and safety systems | Construction | 10 | Not sure | ≈2.5 Billions | Waterfall |

# Appendix II – Interview Code Relationships

This appendix contains a summary of the categories of the answers that were provided by each respondent in the interview-based empirical study presented in this paper. The discussion and the findings are based on the data provided in this appendix. By providing the following tables, the reader will be able to verify the discussion and the findings of the study and assess whether there are other potential relationships that are not related to the research question addressed. The first column shows the respondent's code and the subsequent columns show the coded categories (introduced and detailed in Section 4.2, 4.3 and 4.4) that each respondent mentioned.

Abbreviations used in the table:

- BA: analyst (Business Analyst)
- BT: Business Team
- CC: Customer of customer
- CCS: Challenges related to Commu-nicating and Sharing a unified view
- CEP: Challenges related to the Elicitation Process
- CI: Consultant (Internal)
- CoS: Consultant (Specialist)
- CP: Customer Proxy (invited customer)
- CPE: Challenges related to Predict the Evolution of the system
- CR: Customer (Requester)
- CS: Challenges related to Stakeholders
- CSR: Challenges related to Stable Requirements
- DK: Do not Know
- DST: Developer / development team / Scrum Team
- FO: Function Owner
- GIT: Group Interaction Techniques
- GT: Generic (Technician / Technical team)
- ID: Interaction Designer
- IPT: Isolation Participation Techniques
- MS: Market Research
- MU: Market Unit
- NC: No Challenge
- OET: Other Elicitation Techniques
- ONS: Organization (Not Specified)
- PEU: Potential End-User
- PM: Project Manager
- PO: Product Owner
- REU: Real End-User
- RBS: Reading-Based Techniques
- SA: Software Architect
- SLE: System/Lead Engineer
- SP: Service Provider (carrier)
- TSM: Technical/System Manager
- WE: Web Editor

|  | *RQ1* | *RQ2* | *RQ3* |
|---|---|---|---|
| *S1* | GIT, IPT | BA, CR, DST | NC |
| *S2* | GIT | BA, CR, DST, WE | CEP, CSR, CS |
| *S3* | GIT, IPT, RBS, OET | CP, PEU, | CEP |
| *S4* | GIT | BA, CR, ONS | NC |
| *S5* | GIT, RBS | CR, ONS, SP | CS |
| *S6* | GIT, IPT | CR, MU, TSM | CSR, CS |
| *S7* | DK | CR, ONS | CCS |
| *S8* | GIT, IPT | CC, CR, ONS | CEP, CS |
| *S9* | GIT, IPT | BT, CI, CR | CCS, CPE, CS |
| *S10* | GIT, MS | MU, REU | CCS, CSR |
| *S11* | GIT, IPT | CI, GT, ID, REU | CS |
| *S12* | GIT, IPT | BA, CI, MU, SP | CSR, CS |
| *S13* | IPT, RBS | BA, CR, SA | CCS, CEP, CSR |
| *S14* | GIT, RBS | MU, SLE | CSR |
| *S15* | IPT | CR, PM | CSR, CS |
| *S16* | GIT, MS | DST, ONS | CSR |
| *S17* | IPT, RBS | CR, ONS | CS |
| *S18* | GIT, MS | FO, GT | CSR, CS |
| *S19* | GIT, OET | GT, SLE | CSR, CS |
| *S20* | GIT, OET | ONS | CEP, CS |
| *S21* | GIT, MS | CR, ONS, PO | CEP, CSR, CS |
| *S22* | GIT, RBS | CR, DST | CCS, CSR |
| *S23* | GIT, RBS | BA, CoS, CR, PEU, | CPE |
| *S24* | GIT, OET | BA, CoS, CR, GT, TSM | CSR |

# Appendix III – Relevant statistical correlation values

This appendix contains the relevant correlations that were found in our statistical analysis. For each correlation, we show the p-value and the Cramer's V value. The correlations are organized by RQ.

| RQ | Correlation item 1 | Correlation item 2 | p-value | Cramer's V value |
|---|---|---|---|---|
| RQ1 | Project managers | Group interaction techniques | 0.022 | 0.466 |
| RQ1 | Software architects | Group interaction techniques | 0.022 | 0.466 |
| RQ1 | Consultants | Individual participation techniques | 0.028 | 0.447 |
| RQ1 | Product owners | Market research | 0.022 | 0.466 |
| RQ1 | Function owners | Market research | 0.022 | 0.466 |
| RQ1 | Challenge of instability of requirements | Market research | 0.044 | 0.411 |
| RQ1 | Project number of employees | Individual participation techniques | 0.038 | 0.592 |
| RQ1 | Project costs | Other elicitation techniques | 0.017 | 0.650 |
| RQ2 | Subjects with requirements-related job position | Involvement of customers in elicitation process | 0.016 | 0.655 |
| RQ2 | Projects costs | Involvement of external consultants in elicitation process | 0.012 | 0.674 |
| RQ2 | Project domain | Involvement of external consultants in elicitation process | 0.000 | 1.000 |
| RQ2 | Project methodology | Involvement of real end-users in elicitation process | 0.037 | 0.426 |
| RQ2 | Years working in the organization | Involvement of generic technical roles in elicitation process | 0.022 | 0.466 |
| RQ2 | Years working in their current position | Involvement of generic technical roles in elicitation process | 0.035 | 0.529 |
| RQ2 | Years working in the organization | Involvement of the organization in elicitation process | 0.044 | 0.581 |
| RQ2 | Involvement of analysts in elicitation process | Involvement of external consultants in elicitation process | 0.021 | 0.470 |
| RQ2 | Challenges related to stable requirements | Involvement of market units in elicitation process | 0.044 | 0.411 |
| RQ2 | Challenges related to stable requirements | Involvement of the organization in elicitation process | 0.043 | 0.414 |
| RQ3 | Highest educational background | Challenges related to predict the evolution of the system | 0.028 | 0.723 |
| RQ3 | Years working in the organization | Challenges related to the elicitation process | 0.029 | 0.613 |
| RQ3 | Project domain | Challenges related to communicating and sharing a unified view | 0.010 | 0.793 |
| RQ3 | Project methodology | Challenges related to stakeholders | 0.043 | 0.414 |